\documentclass[aps,prb,reprint,superscriptaddress,floatfix]{revtex4-2}

\usepackage{amssymb}
\usepackage{graphicx}   
\usepackage{color}      
\usepackage{hyperref}   
\usepackage{dcolumn}
\usepackage{orcidlink}
\usepackage{csquotes}
\usepackage{ulem}

\def\lsco{La$_{2-x}$Sr$_x$CuO$_4$}
\def\lbco{La$_{2-x}$Ba$_x$CuO$_4$}
\def\lbcoate{La$_{1.875}$Ba$_{0.125}$CuO$_4$}
\def\lnsco{La$_{1.6-x}$Nd$_{0.4}$Sr$_x$CuO$_4$}
\def\lesco{La$_{1.8-x}$Eu$_{0.2}$Sr$_x$CuO$_4$}

\def\lco{La$_2$CuO$_4$}

\newcommand{\bQ}{\textbf{\textit{Q}}}
\def\newr{\color{black}}
\def\newrr{\color{black}}
\def\newrrr{\color{black}}
\def\newb{\color{black}}

\begin{document}

\title{Charge density waves and pinning by lattice anisotropy in 214 cuprates}

\author{Xiao Hu\orcidlink{0000-0002-0086-3127}}
\affiliation{Condensed Matter Physics and Materials Science Division, Brookhaven National Laboratory, Upton, New York 11973-5000, USA}
\author{P. M. Lozano\orcidlink{0000-0001-6602-4118}}
\altaffiliation{Current address: Advanced Photon Source, Argonne National Laboratory, Argonne, Illinois 60439, USA}
\affiliation{Condensed Matter Physics and Materials Science Division, Brookhaven National Laboratory, Upton, New York 11973-5000, USA}
\affiliation{Department of Physics and Astronomy, Stony Brook University, Stony Brook, New York 11794-3800, USA}
\author{Feng Ye\orcidlink{0000-0001-7477-4648}}
\affiliation{Neutron Scattering Division, Oak Ridge National Laboratory, Oak Ridge, Tennessee 37831, USA}
\author{Qiang Li}
\affiliation{Condensed Matter Physics and Materials Science Division, Brookhaven National Laboratory, Upton, New York 11973-5000, USA}
\affiliation{Department of Physics and Astronomy, Stony Brook University, Stony Brook, New York 11794-3800, USA}
\author{J. Sears\orcidlink{0000-0001-6524-8953}}
\author{I. A. Zaliznyak\orcidlink{0000-0002-8548-7924}}

\author{G. D. Gu\orcidlink{0000-0002-9886-3255}}
\author{J. M. Tranquada\orcidlink{0000-0003-4984-8857}}
\altaffiliation{Contact author: jtran@bnl.gov}
\affiliation{Condensed Matter Physics and Materials Science Division, Brookhaven National Laboratory, Upton, New York 11973-5000, USA}

\date{\today} 

\begin{abstract}
The detection of static charge density waves (CDWs) in \lsco\ (LSCO) with $x\sim0.12$ at relatively high temperatures has raised the question of what lattice feature pins the CDWs.  Some recent structural studies have concluded that some form of monoclinic distortion, indicated by the appearance of certain weak Bragg peaks (type M peaks) at otherwise forbidden positions, are responsible for CDW pinning.  As a test of this idea, we present neutron diffraction results for a single crystal of \lbco\ (LBCO) with $x=1/8$, which is known to undergo two structural transitions on cooling, from high-temperature tetragonal (HTT) to low-temperature orthorhombic (LTO) near 240~K, involving a collective tilt pattern of the corner-sharing CuO$_6$ octahedra, and from LTO to low-temperature tetragonal (LTT) near 56~K, involving a new tilt pattern and the appearance of intensity at peaks of type T.  We observe both type M and type T peaks in the LTT phase, while the type M peaks (but not type T) are still present in the LTO phase.  Given that CDW order is observed only in the LTT phase of LBCO, it is apparent that the in-plane Cu-O bond anisotropy associated with the octahedral tilt pattern is responsible for charge pinning.  We point out that evidence for a similar, but weaker, bond anisotropy has been observed previously in LSCO and should be responsible for CDW pinning there.  In the case of LBCO, the monoclinic distortion may help to explain previously-reported magneto-optical evidence for gyrotropic order.
\end{abstract}

\maketitle

\section{Introduction}

Charge density wave (CDW) order remains a topic of considerable interest, as various types of charge order have now been observed in a variety of quantum materials, including kagome-lattice superconductors $A$V$_3$Sb$_5$ ($A = $ K, Rb, Cs) \cite{orti20,neup22,jian22b,xing24} and antiferromagnet FeGe \cite{teng22}, as well as trilayer nickelate La$_4$Ni$_3$O$_{10}$ \cite{zhan20}, which shows hints of superconductivity under pressure \cite{li24,zhu24,zhan24}, and the insulating reduced version, La$_4$Ni$_3$O$_8$ \cite{zhan16}.  Of course, there is also a significant history of CDW order in cuprate superconductors \cite{hayd24,prou19,comi16}.

In the case of 214 cuprates such as \lbco\ (LBCO), it is clear that the observed charge order \cite{fuji04,huck11} is driven by competition between the kinetic energy of doped holes and the superexchange coupling between Cu spins \cite{emer99,frad15}; nevertheless, the lattice is important for pinning the resulting charge and spin stripes.  The stripes develop within the CuO$_2$ planes, oriented along a Cu-O bond direction.  While there is evidence of dynamic stripes at elevated temperatures \cite{fuji04,miao17,miao19}, static charge stripe order is only observed to develop at or below a structural transition that results in inequivalent Cu-O bonds in orthogonal directions.  This can involve either the low-temperature tetragonal (LTT) or the low-temperature less-orthorhombic (LTLO) phase \cite{craw91,huck11,bozi15}.  A similar relationship between charge order and lattice symmetry is observed in \lnsco\ \cite{ichi00,gupt21} and \lesco\ \cite{fink11,lee22}.  The reduction in lattice symmetry is evident in diffraction from the appearance of finite intensity at reciprocal lattice points where signal is forbidden in the higher-symmetry phase.  For the LTT and LTLO phases, we label the new reflections common to both as type T.

A challenge to this stripe pinning picture came with the discovery of charge order in \lsco\ \cite{wu12,crof14,chri14,tham14}, where the structure is low-temperature orthorhombic (LTO) \cite{rada94}.  In the LTO phase, the CuO$_6$ octahedra tilt along a diagonal direction with respect to the in-plane Cu-O bonds, with the consequence that the approximately orthogonal in-plane Cu-O bonds are equivalent.  What could cause the pinning of charge stripes?

One possibility appeared with the {\newr surprising} discovery, in the parent compound \lco, of a set of weak diffraction peaks (which we label type M) that are forbidden under the previously assumed symmetry (space group $Bmab$) of the LTO phase \cite{reeh06}.  Similar single-crystal diffraction studies eventually demonstrated the appearance of the same peaks in \lsco\ (LSCO) $x=0.05$ \cite{sing16} and $x=0.12$ \cite{fris22}.  To allow for them, a low symmetry monoclinic space group is required, although no monoclinic distortion of the unit cell has been detected.  {\newr (Note that several extensive powder diffraction studies covering the LSCO \cite{rada94} and LBCO \cite{axe89,suzu89a,kata93} phase diagrams never identified a unit-cell symmetry other than tetragonal or orthorhombic. This is why the discovery of the lower-symmetry reflections \cite{reeh06} was so surprising.)} 
With the reduced symmetry, there are many more parameters required to describe the positions of atoms within the unit cell.  In fitting the measured diffraction intensities, constraints have been applied to the possible position parameters.  The most significant features involve broken symmetry between nearest-neighbor CuO$_6$ octahedra, primarily with respect to the apical oxygens. Models have involved unequal tilts of the apical oxygens on the same octahedron \cite{reeh06} or on neighboring octahedra \cite{sing16}.  In their study of LSCO $x=0.12$, Frison {\it et al.} \cite{fris22} concluded that the monoclinic lattice distortions provide the appropriate symmetry reduction required for the charge stripe order.  A recent x-ray coherent scattering study of an LSCO $x=0.12$ crystal attempted to associate the charge order with the thermal development of type M peaks \cite{shen23}.

In this paper, we present a neutron diffraction study on a single crystal of LBCO $x=0.125$.  We observe the presence of the type M diffraction peaks in the LTT phase; however, we find that they are also present, with little change, in the LTO phase.  In contrast, the intensities of the type T peaks unique to the LTT octahedral tilt pattern, show an abrupt drop at the LTT-to-LTO transition.  We conclude that the monoclinic distortions are not relevant to the pinning of charge stripes in LBCO.  For the case of LSCO, we point out previous observations of type T peaks in $x=0$ \cite{sapk21}, $x=0.07$ \cite{jaco15}, and $x=0.12$ \cite{fris22}, which indicate that the structure there corresponds to LTLO rather than LTO.  The associated broken symmetry between in-plane Cu-O bonds, though small, should allow some pinning of charge stripes.


The monoclinic distortions, while not key to stripe pinning, may be relevant to understanding the gyrotropic response \cite{hosu13} observed in magneto-optical studies of LBCO \cite{kara12,kara14}.  We have observed broken symmetry in the LTT phase between the two CuO$_2$ planes within a unit cell, resulting in finite intensity for reflections of type $(0,0,L)$ with $L$ odd, and the space group we have used to fit the diffraction data in the LTT phase, $P2_111$, 
is compatible with a chiral crystal structure.

The rest of the paper is organized as follows. We describe our experimental methods in the following section and then present the data and its analysis in Sec.~\ref{sc:dat}.   The significance of the results is discussed in Sec.~\ref{sc:disc}, and we end with a summary of our conclusions in Sec.~\ref{sc:conc}.

\begin{figure*}[t]
 \centering
    \includegraphics[width=1.7\columnwidth]{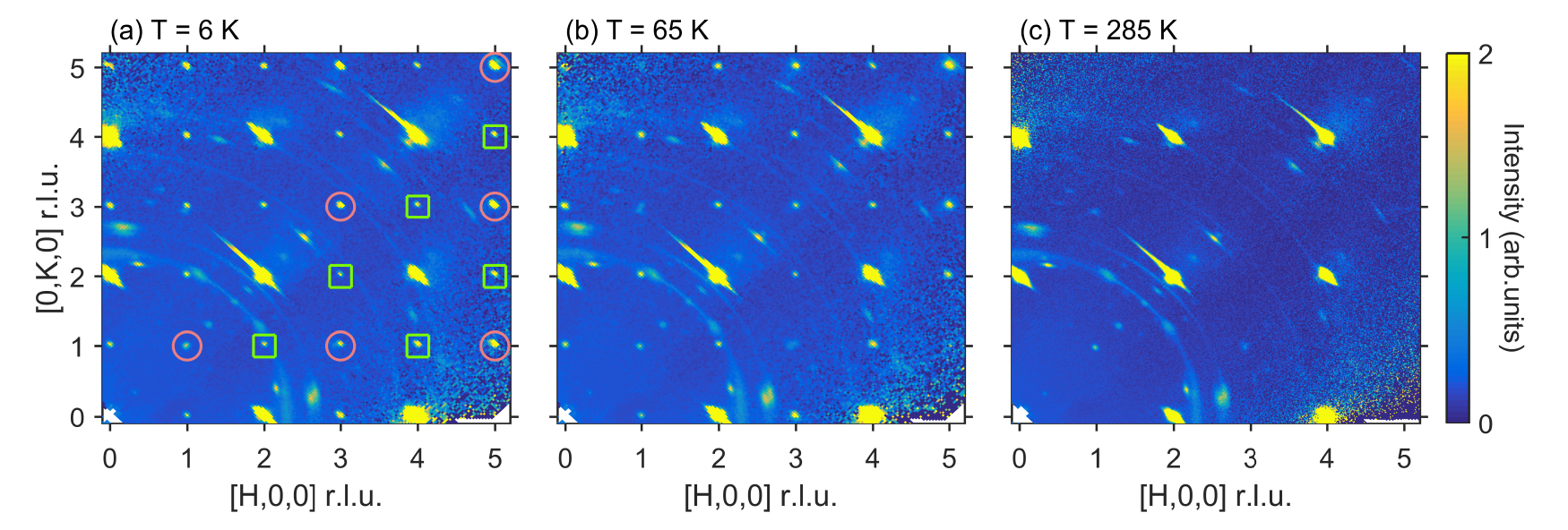}
    \caption{Contour plots of neutron diffraction intensities for \lbcoate\ measured at (a) 6, (b) 65, and (c) 285 K in the $(H,K,0)$ scattering plane using the lattice parameters $a=b=5.35(1)$~\AA, $c=13.2(1)$~\AA. The measured intensity was averaged over $L\in [-0.1,0.1]$. {\newr Representative T-type and M-type reflections are highlighted in panel (a) by red circles and green squares, respectively. Extrinsic contributions (aluminum, La$_4$BaCu$_5$O$_{13+\delta}$ impurity phase) are manifested as weak signals apart from the bright Bragg peaks in (c); these are also present in (a) and (b).}}
    \label{fg:contour}
\end{figure*}

\section{Experimental Methods}

The sample was obtained from a crystalline rod grown at Brookhaven by the traveling-solvent floating-zone method using a recipe first developed by Fujita and coworkers at the Institute of Materials Science, Tohoku University \cite{fuji04}.  Throughout the paper, we will discuss the crystal lattice in terms of the LTT unit cell, $a=b=5.35(1)$~\AA, $c=13.2(1)$~\AA, and the momentum transfer, \bQ\ $= (H,K,L)$, in the corresponding reciprocal lattice units (r.l.u.) of $(2\pi/a,2\pi/b,2\pi/c)$.  A piece of crystal was oriented by Laue diffraction and cut to the shape of a cube of size $\approx 3 \times 3 \times 3$~mm$^3$, with faces perpendicular to each of the unit cell axes.  Magnetic susceptibility measurements as a function of temperature with field applied parallel to the $c$ axis confirmed that the sample was consistent with previous characterizations of LBCO $x=1/8$ \cite{tran08}.

The single-crystal neutron diffraction experiment was performed at CORELLI \cite{ye18}, Beamline 9 at the Spallation Neutron Source, Oak Ridge National Laboratory.  To provide an efficient measurement of elastic diffuse scattering, the instrument combines an incident white beam and a statistical chopper.  It is able to simultaneously collect scattering using a large number of neutron energies; however, the energy resolution also varies with energy, which is a slight complication.

At CORELLI, the LBCO $x=1/8$ single crystal was glued to an aluminum pin and mounted in a closed cycle refrigerator with the $b$ axis vertical, so that the $(H,0,L)$ reciprocal lattice zone was in the horizontal scattering plane.  Note that the crystal is twinned, so that both $(H,0,L)$ and $(0,K,L)$ planes were represented in the horizontal.  Measurements were carried out by rotating the sample about the vertical axis in 1$^{\circ}$ steps over a 360$^{\circ}$ range. A counting time of $\sim$210~s was used for each step, giving a total measurement time of $\sim$23 hours for each of the data sets, which were collected at 6, 65, and 285~K.  The data reduction and binning to a rectangular grid were performed using MantidWorkbench \cite{mantid14}. Elastic scattering is achieved via energy discrimination by implementing the cross-correlation technique \cite{ye18}. The crystal structure information was obtained from refinements using the FullProf algorithm \cite{FullProf93}.  In addition, temperature dependence was measured for a fixed orientation of the sample, providing more detailed information for a limited set of reflections.

\section{Data and Analysis}
\label{sc:dat}

\begin{figure}[b]
 \centering
    \includegraphics[width=0.99\columnwidth]{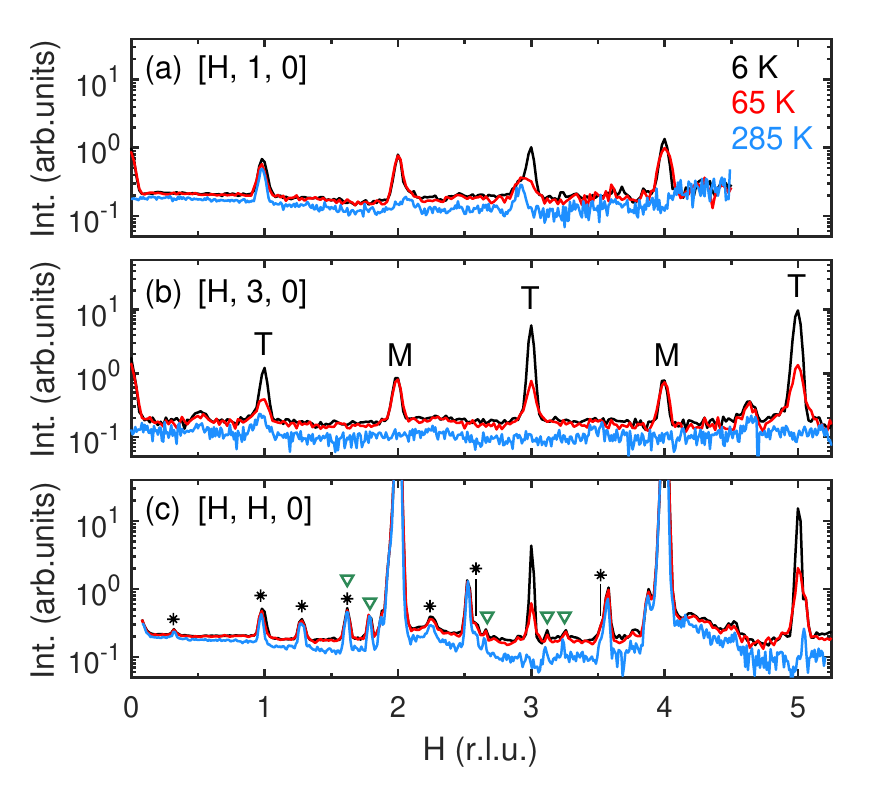}
    \caption{Intensity vs.\ {\bf Q} for cuts along (a) $[H,1,0]$, (b) $[H,3,0]$, and (c) $[H,H,0]$ measured at temperatures of 6~K (black), 65~K (red), and 285~K (blue). In (c), $*$ marks peaks associated with the impurity La$_4$BaCu$_5$O$_{13+\delta}$ phase \cite{mich87}, while  $\triangledown$ denotes Al peaks from the sample environment. {\newr Representative T-type and M-type peaks as mentioned in the main text are labeled \enquote{T} and \enquote{M}, respectively, in (b).}}
    \label{fg:cuts}
\end{figure}

Figure~\ref{fg:contour} presents color contour maps of the neutron scattering patterns we obtained for the $(H,K,0)$ zone; note that the maximum intensity is set to a low value to make weak features distinguishable. At $T= 285$~K, shown in Fig.~\ref{fg:contour}(c), the only diffraction peaks occur for $H$ and $K$ even; their plotted intensities are saturated, as they extend three-orders of magnitude beyond the selected scale.  The very weak sharp features that remain at this temperature are associated with low-concentration impurity phases such as La$_4$BaCu$_5$O$_{13+\delta}$ \cite{mich87} and scattering from the Al sample mount, as indicated in Fig.~\ref{fg:cuts}(c).

At 6 K, where the sample has transformed to the LTT phase, Fig.~\ref{fg:contour}(a) shows that intensity has appeared at every reciprocal lattice point where $H$ and $K$ are integer.  Within the $(H,K,0)$ zone, the T-type peaks known to be associated with the LTT phase correspond to those where $H$ and $K$ are both odd, such as (3,3,0).  The peaks for which $H$ and $K$ form an odd-even or even-odd combination, such as (1,4,0), are of the M type.  The early crystallographic studies of LBCO \cite{axe89,axe89b} associated the LTT phase with space group $P4_{2}/ncm$ (No.~138), for which the type T peaks are allowed; this was based on measurements by powder diffraction, which lacks the dynamic range to detect the M-type peaks.

To appreciate the situation at 65~K, it is helpful to look at Fig.~\ref{fg:cuts}.  The intensities of type T peaks, such as (3,3,0) and (5,3,0) in Fig.~\ref{fg:cuts}(b), have dropped by an order of magnitude, while the type M peaks, such as (2,3,0) and (4,3,0), show little change.  As the LTT to LTO transition is known to be of first order \cite{wen12a}, the small but finite signal of type T peaks can be attributed to residual LTT phase \cite{bozi15}.

\begin{figure}[b]
 \centering
    \includegraphics[width=0.95\columnwidth]{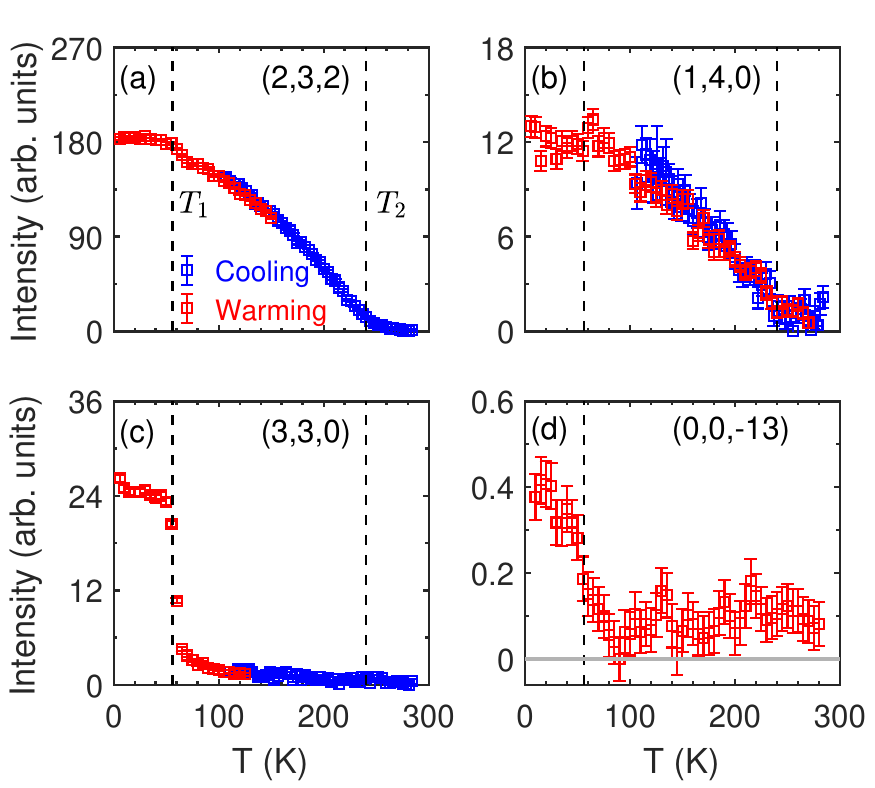}
    \caption{Temperature-dependent integrated peak intensities {\newrrr (integrated in one dimension, along $Q$)} of LTO/LTT peak $(2,3,2)$ (a), type-M peak $(1,4,0)$ (b), type-T peak $(3,3,0)$ (c), and peak $(0,0,-13)$ (d). The vertical dashed lines represent the transition temperatures 56 K ($T_1$) and 240 K ($T_2$), respectively.}
    \label{fg:tdep}
\end{figure}

Before considering structural refinements, we look at the temperature dependence of representative reflections, as shown in Fig.~\ref{fg:tdep}.  As already noted, LBCO $x=0.125$ has two structural transitions on warming, from LTT to LTO at $T_1 \approx 56$~K and from LTO to HTT at $T_2 \approx 240$~K \cite{huck11}.  The tilts of the CuO$_6$ octahedra in both the LTO and LTT phases result in finite intensity at $(H,K,L) = ({\rm even},{\rm odd},{\rm even})$ with $L\neq 0$, with $(H,K,L)$ and $(K,H,L)$ being equivalent in the LTT phase. 
An example is (2,3,2), shown in Fig.~\ref{fg:tdep}(a), where the intensity onsets near $T_2$, grows on cooling, and shows just a small rise at $T_1$.  We find that the M-type reflection (1,4,0) shows a very similar behavior, as seen in Fig.~\ref{fg:tdep}(b).  In contrast, the T-type reflection (3,3,0) develops abruptly at $T_1$; the tail seen in Fig.~\ref{fg:tdep}(c) just above $T_1$ is due to residual LTT domains, as mentioned above.
We also observed a small but finite intensity at $(0,0,L)$ reflections with $L$ odd in the LTT phase, as shown in Fig.~\ref{fg:tdep}(d), which indicates the breaking of the screw symmetry along the $c$ axis previously reported \cite{axe89} in the LTT phase.

\begin{table*}
	\caption{Results of structural refinement of 4801 single-crystal diffraction peak intensities measured with neutron wavelengths 0.81--1.28~\AA\ (1549 unique $HKL$ values) for \lbco\ with $x=0.125$ at $T=6$~K. Figures in parentheses are estimated standard deviations in the last digit. The space group is $P2_111$.  The lattice parameters are $a=5.3549$~\AA, $b=5.3553$~\AA, $c=13.1949$~\AA, with $\alpha=\beta=\gamma=90^\circ$. ``La'' represents 0.9375 occupancy of La and 0.0625 occupancy of Ba. \label{6K}}.
	\begin{ruledtabular} 
		\begin{tabular}{ccdddd}
  & & \multicolumn{3}{c}{\quad\quad$(0,0,\frac34)\quad +$}\vspace{3pt} & \\ 
   \cline{3-5} 
 Atom & Site & \multicolumn{1}{c}{$x$\rule{0pt}{9pt}} & \multicolumn{1}{c}{$y$} & \multicolumn{1}{c}{$z$} & \multicolumn{1}{r}{$U_{\rm iso}$ (\AA$^2$)} \\
\hline
  Cu(1)\rule{0pt}{9pt} & $2a$ & 0.0 & 0.0 & 0.0 & 0.0022(1) \\
  O(1) & $2a$ & 0.25 & 0.25 & 0.0076(1) & 0.0073(1)  \\
  O(2) & $2a$ & -0.25 & 0.25 & 0.0 & 0.0073(1) \\
  O(3) & $2a$ & -0.0048(5) & -0.0150(6) & 0.1823(1) & 0.0073(1) \\
  O(4) & $2a$ & 0.0325(5) & 0.0143(4) & -0.1828(1) & 0.0073(1) \\
  La(1) & $2a$ & -0.0008(3) & 0.0052(4) & 0.3614(1) & 0.0012(1) \\
  La(2) & $2a$ & -0.0094(3) & -0.0040(3) & -0.3602(1) & 0.0012(1) \\
\hline\hline
  & & \multicolumn{3}{c}{\quad\quad$(\frac12,\frac12,\frac34)\quad +$\rule{0pt}{10pt}}\vspace{3pt} & \\ 
   \cline{3-5} 
 Atom & Site & \multicolumn{1}{c}{$x$\rule{0pt}{9pt}} & \multicolumn{1}{c}{$y$} & \multicolumn{1}{c}{$z$} & \multicolumn{1}{r}{$U_{\rm iso}$ (\AA$^2$)} \\ 
  \hline
  Cu(2) & $2a$ & 0.0 & 0.0 & 0.0 & 0.0022(1)\\
 O(5) & $2a$ & 0.25 & 0.25 & -0.0076(1) & 0.0073(1) \\
 O(6) & $2a$ & -0.25 & 0.25 & 0.0 & 0.0073(1) \\
 O(7) & $2a$ & 0.0270(5) & 0.0117(6) & 0.1818(1) & 0.0073(1) \\
 O(8) & $2a$ & -0.0209(5) & -0.0163(4) & -0.1837(1) & 0.0073(1) \\
 La(3) & $2a$ & -0.0090(3) & -0.0030(3) & 0.3600(1) & 0.0012(1) \\
 La(4) & $2a$ & -0.0007(3) & 0.0031(4) & -0.3616(1) & 0.0012(1) \\
		\end{tabular}
	\end{ruledtabular}
 Extinction$=15.9(5)$, $\chi^2 = 28.4$, $R_F = 5.3$\% \rule{0pt}{9pt}
\end{table*}

Charge and spin stripe order in LBCO $x=0.125$ are observed only below $T_1$ \cite{fuji04,huck11,wilk11}.  Based on the results in Fig.~\ref{fg:tdep}, it seems reasonable to conclude that the broken rotational symmetry of in-plane Cu-O bonds, indicated by type T reflections, is essential to stripe pinning, while the distortions associated with type M peaks are not (or at least are not sufficient).

In order to fit the data to a structural model, we have to choose an appropriate space group for the structure.  The original choice for the LTT phase, space group $P4_2/ncm$ 
\cite{axe89b}, which captures the rigid tilts of the CuO$_6$ octahedra that lead to the type T peaks, does not include type M peaks.  To account for the type M peaks that were observed in orthorhombic \lco, Reehuis {\it et al.} \cite{reeh06} used monoclinic space group $Bm11$ (No.~8) which allows for distortions of the octahedra through displacements of the apical oxygens. For LSCO with $x=0.05$, Singh {\it et al.} \cite{sing16} made use of $B2/m11$ (No.~12), which allows for a slightly different set of apical oxgyen displacements.  Neither of these monoclinic space groups allows for the type T reflections.  

\begin{figure}[b]
 \centering
    \includegraphics[width=1.0\columnwidth]{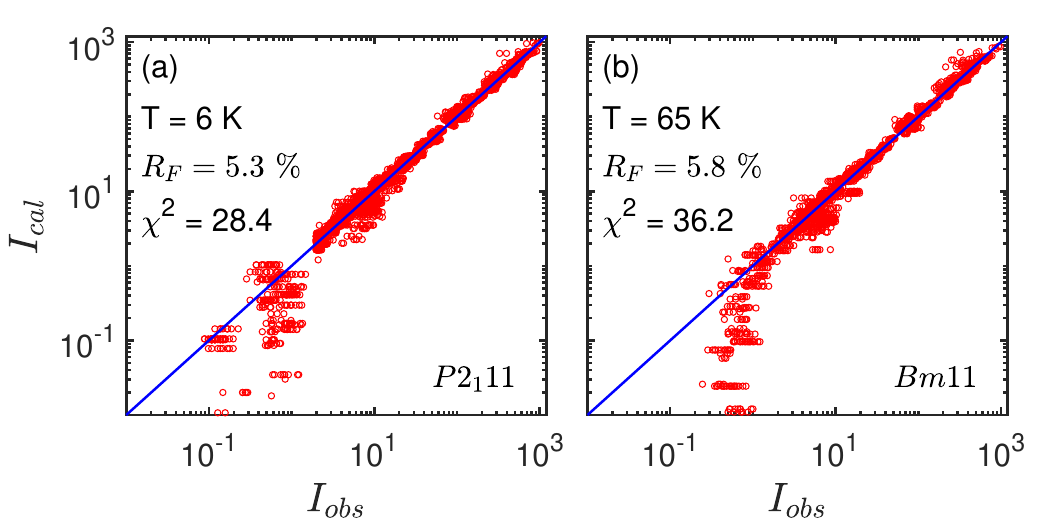}
    \caption{Refinement results of the neutron diffraction data, comparing calculated and observed intensities, at $T = 6$ K with the monoclinic $P2_111$ space group (a) and at $T=65$ K with the monoclinic $Bm11$ space group (b).}
    \label{fg:datavsfit}
\end{figure}

\begin{figure}[b]
\centering
    \includegraphics[width=0.95\columnwidth]{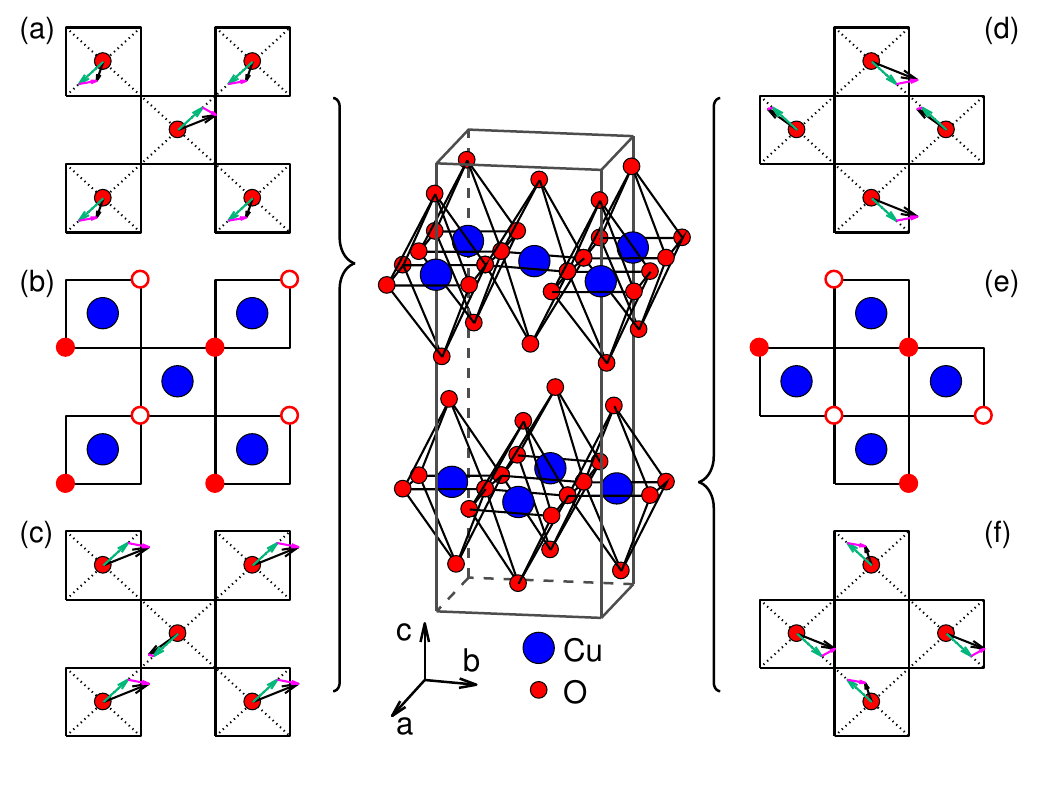}
    \caption{Schematic diagram of crystal structure obtained from the refinement of 6 K neutron diffraction data with the monoclinic $P2_111$ space group, with Cu in blue and O in red. Side panels represent the atomic configuration in each layer of octahedra: (a)$-$(c) upper layer; (d)$-$(f) lower layer. Black arrows indicate exaggerated horizontal atomic displacements of apical oxygens; these are a combination of the LTT-like tilt (green arrows) and the monoclinic distortion (magenta arrows). (a),(d) and (c),(f) describe apical oxygen configurations above and below the corresponding Cu-O planes, respectively. Hollow/filled red circles in (b) and (e) indicate tilted in-plane oxygen atoms above/below the plane.}
    \label{fg:ltt_p21str}
\end{figure}

As noted previously \cite{sapk21}, to be able to describe both the type T and type M peaks it is necessary to choose a primitive monoclinic space group.  To fit the diffraction data, we use $P2_111$ (No.~4).  The parameters obtained from the refinement are presented in Table~\ref{6K} and the observed and calculated intensities are compared in Fig.~\ref{fg:datavsfit}(a).  The type M reflections tend to have relative intensities on the scale of $10^{-4}$--$10^{-3}$ with respect to strong Bragg peaks, and the calculated intensities tend to be smaller than the measured intensities in this regime.  For these weak reflections, one must also consider the possibility of contamination by multiple scattering, whose probability grows as the neutron wavelength gets shorter. 
While the fitting is imperfect, we are most interested in the qualitative information that we can extract.

To minimize the number of variables, we constrained the in-plane Cu and O atoms to positions consistent with $P4_2/ncm$; the distortions allowed by the reduced symmetry are then realized by the apical O and the La sites.  To illustrate these distortions, the displacements of the apical oxygens are indicated by black arrows in Fig.~\ref{fg:ltt_p21str}.  (Relaxing our constraints yields marginal reductions of $\chi^2$ and no additional insight.)

To obtain a clearer picture, we decompose the apical O positions into distinct components.  The displacements that contribute to the type T reflections correspond to those of Wyckoff site $8(i)$ for space group $P4_2/ncm$, with $|x|=|y|=0.0175$ and $|z|=0.1824$, indicated by green arrows in Fig.~\ref{fg:ltt_p21str}.  Relative to those positions, the monoclinic distortions for the different sites correspond to
\begin{eqnarray}
    {\bf\Delta}_{\rm O(3)} & = & ( 0.0127,     0.0025,  -0.0001), \\
    {\bf\Delta}_{\rm O(4)} & = & (  0.0150,   -0.0032,  -0.0004), \\
    {\bf\Delta}_{\rm O(7)} & = & (  0.0095,  -0.0058,  -0.0006), \\
    {\bf\Delta}_{\rm O(8)} & = & ( -0.0035,   0.0012,  -0.0013). 
\end{eqnarray}
These can be characterized as dipolar distortions of the apical oxygens relative to the Cu, predominantly parallel to the $a$-$b$ plane; they are indicated by magenta arrows in Fig.~\ref{fg:ltt_p21str}.

{\newb As we noted above, space group $P2_111$ allows intensity at $(0,0,L)$ reflections with $L$ odd.  The algorithm that evaluated the integrated intensities of all diffraction peaks within the measured range included signal for $(0,0,-3)$ but did not identify a peak at $(0,0,-13)${\newrr; the temperature-dependent intensity indicated in Fig.~\ref{fg:tdep}(d) was obtained by integrating the peak intensity at $(0,0,-13)$ in one dimension}.  In any case, the fitted model does yield finite intensity for these peaks, and for $(0,0,-13)$ the calculated {\newrr integrated} intensity is comparable to the observed one {\newrr when one takes account of typical peak widths}.
}

Next, we consider the results at $T=65$~K.  
{\newb The temperature dependence of the $(3,3,0)$ peak shown in Fig.~\ref{fg:tdep}(c) drops abruptly at 56~K, consistent with a first-order transition \cite{huck11}. For such a transition, it is possible to have both the LTT and LTO phases coexisting over a finite range of temperature, as was clearly observed in the case of LBCO $x=0.095$ \cite{wen12a}.  In a previous study of LBCO $x=0.125$ on the HYSPEC instrument, finite elastic intensity for $(3,3,0)$ was observed at 60~K but not at temperatures of 100~K or higher \cite{bozi15}.  This suggests that the intensity just above the transition corresponds to a small volume fraction of LTT domains that coexist with the dominant LTO phase.  On HYSPEC, we also observed gapless soft phonons centered at $(3,3,0)$ for $T=60$~K and above \cite{bozi15}.  In light of those results, we believe that the tail of $(3,3,0)$ intensity seen in Fig.~\ref{fg:tdep}(c) is a consequence of residual LTT domains. 
As a result, we removed type T peaks from the refinements at 65~K. }

\begin{table*}
	\caption{Results of structural refinement of 5864 single-crystal diffraction peak intensities with neutron wavelengths 0.81--1.28~\AA\ (1477 unique $HKL$ values) for \lbco\ with $x=0.125$ at $T=65$~K. Figures in parentheses are estimated standard deviations in the last digit. The space group is $Bm11$.  The lattice parameters are $a=5.3572$~\AA, $b=5.3595$~\AA, $c=13.1976$~\AA, with $\alpha=\beta=\gamma=90^\circ$. ``La'' represents 0.9375 occupancy of La and 0.0625 occupancy of Ba. \label{65K}}
	\begin{ruledtabular} 
		\begin{tabular}{ccdddd}
  & & \multicolumn{3}{c}{\quad\quad$(0,0,0)\quad +$}\vspace{3pt} & \\ 
   \cline{3-5} 
 Atom & Site & \multicolumn{1}{c}{$x$\rule{0pt}{9pt}} & \multicolumn{1}{c}{$y$} & \multicolumn{1}{c}{$z$} & \multicolumn{1}{r}{$U_{\rm iso}$ (\AA$^2$)} \\
\hline
  Cu(1)\rule{0pt}{9pt} & $2a$ & 0.0 & 0.0 & 0.0 & 0.0032(1) \\
  O(1) & $4b$ & 0.25 & 0.25 & 0.0052(1) & 0.0082(1)  \\
  O(2) & $2a$ & 0.0 & -0.0256(4) & 0.1788(1) & 0.0082(1) \\
  O(3) & $2a$ & 0.0 & 0.0243(4) & -0.1864(1) & 0.0082(1) \\
  La(1) & $2a$ & 0.0 & -0.0028(3) & 0.3588(1) & 0.0021(1) \\
  La(2) & $2a$ & 0.0 & -0.0121(2) & -0.3629(1) & 0.0021(1) \\
\hline\hline
  & & \multicolumn{3}{c}{\quad\quad$(\frac12,\frac12,0)\quad +$\rule{0pt}{10pt}}\vspace{3pt} & \\ 
   \cline{3-5} 
 Atom & Site & \multicolumn{1}{c}{$x$\rule{0pt}{9pt}} & \multicolumn{1}{c}{$y$} & \multicolumn{1}{c}{$z$} & \multicolumn{1}{r}{$U_{\rm iso}$ (\AA$^2$)} \\ 
  \hline
 Cu(2) & $2a$ & 0.0 & 0.0 & 0.0 & 0.0032(1)\\
 O(4) & $4b$ & 0.25 & 0.25 & -0.0052(1) & 0.0082(1) \\
 O(5) & $2a$ & 0.0 & 0.0320(4) & 0.1795(1) & 0.0082(1) \\
 O(6) & $2a$ & 0.0 & -0.0172(5) & -0.1854(1) & 0.0082(1) \\
 La(3) & $2a$ & 0.0 & -0.0113(2) & 0.3600(1) & 0.0021(1) \\
 La(4) & $2a$ & 0.0 & 0.0010(3) & -0.3616(1) & 0.0021(1) \\
		\end{tabular}
	\end{ruledtabular}
 Extinction$=9.6(4)$, $\chi^2 = 36.2$, $R_F = 5.8$\% \rule{0pt}{9pt}
\end{table*}

It is clear from Fig.~\ref{fg:tdep}(b) that type M peaks are still present at 65~K and, as already mentioned, these are not described by the $Bmab$ space group originally associated with the LTO phase \cite{gran77,jorg87,axe89}.  In Table~\ref{65K}, we present the results of a refinement with space group $Bm11$, which was introduced to explain the presence of type-M peaks in the LTO phase of \lco\ \cite{reeh06}. In this fit, the monoclinic distortions of the apical O involve a significant parallel shift along the $c$ axis. We find that the details of the distortion depend on the space group used to fit the data, as this can result in different numbers of parameters to model it.  For example, at 65~K we can get a slightly better quality of fit with $P2_111$ ($\chi^2 = 30.5$); in contrast, a fit with $B2/m11$ \cite{sing16} yields a somewhat worse fit ($\chi^2=48.6$).

\section{Discussion}
\label{sc:disc}

{\newb Our association of the CDW order with the T type peaks relies on experimental evidence that the CDW order in LBCO occurs only in the LTT phase.  For $x=0.125$, this is supported by single-crystal neutron diffraction \cite{fuji04} and hard x-ray diffraction \cite{huck11} studies, along with transport measurements that show abrupt changes in the $c$-axis resistivity, in-plane thermopower, and in-plane thermal conductivity at the LTO-LTT transition \cite{tran08}.  To be complete, we need to discuss experimental results that might cause confusion.  One of these is the x-ray diffraction observation that the CDW order survives the LTT-HTT transition induced by hydrostatic pressure at $\sim1.8$ GPa for $T\lesssim50$~K \cite{huck10}.  Here it is important to note that the integrated intensity of the CDW diffraction peak and the associated correlation length are {\newrr both} reduced in the high-pressure HTT phase. Furthermore, superlattice peaks associated with octahedral tilts survive with reduced intensity and an {\newrr in-plane} correlation length {\newrr similar to that of the CDW correlations. Although the drop in intensity of the octahedral tilt peaks is very large on approaching the HTT phase, it involves a long-range order along the $c$ axis, whereas the CDW has a very short correlation length along $c$. In the HTT phase, the octahedral tilts likely have a similar short $c$-axis correlation length. Later measurements to higher pressure found that the peak widths and intensities of the CDW and octahedral-tilt peaks evolve in an identical fashion in the HTT phase, with both disappearing at $\sim3.6$~GPa \cite{fabb16}.  Fits to} x-ray absorption fine structure {\newrr measurements} at the La $K$ edge demonstrated that the splitting of the local La-O distances present in the LTT phase survives into the HTT phase, though with decreasing magnitude \cite{fabb13,fabb16}. {\newrr (To be clear, M-type distortions were not considered in that analysis.)}  Hence, it appears that short-range LTT structural correlations survive in the HTT phase along with the CDW correlations. 

Another possible point of confusion comes from the resonant inelastic x-ray scattering (RIXS) measurements at the Cu $L_3$ edge around the CDW wave vector for temperatures across the LTT-LTO transition by Miao {\it et al.} \cite{miao17}.  Taken at face value, those measurements suggest that a very weak and broadened CDW peak survives into the LTO phase with a shifted ordering wave vector.  To evaluate those results, it is crucial to note that the energy resolution was 70 meV, comparable to the energy of the softened phonon in the Cu-O bond-stretching mode \cite{rezn06,wang21a} and larger than phonon anomalies seen in lower-energy modes \cite{miao19,huan21}. The simplest way to reconcile all results (including the transport measurements mentioned above) is to conclude that the measurements of Miao {\it et al.} \cite{miao17} in the LTO phase represent dynamic CDW correlations measured in a ``quasielastic'' fashion.}

We have seen that stripe ordering in LBCO is associated with broken rotational symmetry of the planar Cu-O bonds, which is indicated by the presence of T-type Bragg intensities, with no unique correlation to the monoclinic distortions.  Is there evidence for type T peaks in LSCO? Indeed, there is.  Peaks such as the (3,3,0) have been detected in \lsco\ with $x=0$ \cite{sapk21}, 0.07 \cite{jaco15}, and 0.12 \cite{fris22}.  

In the case of LSCO, the type T peaks are observed within an orthorhombic phase.  If we ignore the monoclinic distortions and consider the picture of structural phases determined by rotations of rigid octahedra \cite{axe94}, the conventional LTO phase of LSCO, associated with space group $Bmab$, involves rotations of the octahedra around an in-plane axis at 45$^\circ$ to the Cu-O bonds.  The presence of finite intensity for T-type peaks indicates that there is also a tilt component around an in-plane Cu-O bond axis, which breaks the symmetry between Cu-O bonds at 90$^\circ$ to each other.  Such a phase was identified in \lnsco\ (LNSCO) \cite{craw91} and associated with space group $Pccn$. It has also been seen in LBCO for dopant concentrations away from 0.125 \cite{huck11}; it has sometimes been labeled the low-temperature less-orthorhombic (LTLO) phase.  In LNSCO and LBCO, there is a structural transition on cooling from the LTO to the LTLO phase that can be first order \cite{craw91,wen12a}.

In terms of octahedral tilts, we believe that the orthorhombic phase of LSCO should be characterized as LTLO.  While it may need further investigation, there is no evidence for a transition from LTO to LTLO \cite{jaco15,sapk21}.  The Cu-O bond anisotropy associated with the LTLO phase determines a unique orientation for the ordering of charge stripes that is observed for $x\sim0.12$ \cite{crof14,chri14,tham14,wen19,miao21,vona23}.

A related aspect of anisotropy was recently addressed in a study of stripe order in LSCO \cite{he24}.  For $x=0.12$, the stripe orientation is rotated by $3^\circ$ relative to the Cu-O bond direction. The ordering was modelled by numerical analysis of a single-band Hubbard model with both nearest- and next-nearest-neighbor hopping. It was found that a small anisotropy in the next-nearest-neighbor hopping could explain the stripe rotation.  Not discussed there was the anisotropy that selects a unique stripe orientation in the first place.  In the numerical calculations, anisotropic boundary conditions (periodic in one direction, open in the other) play this role.  In the real material, we argue that it is the anisotropy in nearest-neighbor hopping that does the trick.

The bond anisotropy is weaker in the LTLO phase of LSCO than in the LTT phase of LBCO. A consequence of this is that the volume fraction that exhibits stripe order for $x\sim0.12$ is significantly smaller in LSCO than in LBCO.  This is evident from neutron \cite{he24} and x-ray \cite{wen19,wen23} scattering, as well as from muon spin relaxation studies \cite{savi02,savi05} and nuclear magnetic resonance \cite{arse20}.  This means that there is a larger volume fraction of spatially uniform superconductivity in LSCO and a higher bulk superconducting transition compared to LBCO $x=0.125$ \cite{savi05}.

It should be clear that the charge ordering occurs in the presence of appropriate anisotropy, but that structural anisotropy can be present without any charge order. For example, the LTLO phase appears below 76 K in La$_{1.65}$Nd$_{0.35}$CuO$_4$, which is an antiferromagnetic insulator with a N\'eel temperature of 316~K \cite{keim93}.  In the uncorrelated band insulator La$_2$MgO$_4$, an LTO to LTT transition occurs on cooling below 350~K \cite{tide22}.

While the monoclinic distortions do not appear to be relevant for stripe pinning, they nevertheless can have measurable impacts. In particular, we note that $P2_111$, which we find to be necessary to accommodate both the T and M peaks, is a Sohncke group \cite{fech22}, which means that it is compatible with a chiral crystal structure.  This is relevant to explaining magneto-optical measurements on LBCO $x=0.125$ that revealed the onset of a Kerr rotation at the structural transition to the LTT phase \cite{kara12}. This was  initially interpreted as evidence for time-reversal symmetry breaking associated with charge-stripe ordering \cite{kara12}; however,  theoretical analysis \cite{hosu13} and further measurements led to the conclusion that the Kerr rotation must be due to a gyrotropic effect associated with some sort of chiral order \cite{kara14}.  It seems likely that a chiral component of the monoclinic distortions is the answer.

Of course, the distortions of the octahedra modify the crystal field seen by the Cu sites, thus affecting the energy splittings among the $3d$ orbitals \cite{huan23b}, which impacts magnetic exchange terms through spin-orbit coupling \cite{dzya58,mori60}. In LBCO $x=0.125$, an anomalous Nernst effect was observed to set in with the structural transition and the charge order \cite{li11}; again, this was viewed as evidence of time-reversal symmetry breaking.  Perhaps the monoclinic distortions could play a role in finding an explanation for this effect.

\section{Conclusions}
\label{sc:conc}

We have shown that distinct Bragg peaks appear in LBCO $x=0.125$ at low temperature, with one set corresponding to the octahedral tilts that break in-plane Cu-O bond symmetry and another set that correspond, dominantly, to monoclinic distortions of the apical oxygens in the octahedra.  The stripe order occurs only in the low-temperature phase with broken bond symmetry, but the monoclinic distortions remain at higher temperature.  We conclude that the monoclinic distortions are not relevant to the ordering of charge stripes; the key is the anisotropy associated with the broken bond symmetry.  We have pointed out that there is evidence for similar but weaker broken bond symmetry in LSCO, which we presume is the cause of local stripe pinning.

While the monoclinic distortions do not impact the charge order, they may provide an understanding for past observations of Kerr rotation in LBCO.  The space group we used to refine the structure of the LTT phase allows for chiral features, which are needed to explain the optical response.

\section{Acknowledgments}

Work at Brookhaven is supported by the Office of Basic Energy Sciences, Materials Sciences and Engineering Division, U.S. Department of Energy (DOE) under Contract No.\ DE-SC0012704.   A portion of this research used resources at the Spallation Neutron Source, a DOE Office of Science User Facility operated by Oak Ridge National Laboratory. The beam time on CORELLI was allocated under Proposal No. IPTS-27254.

\bibliography{LNO,theory,neutrons}

\end{document}